\documentclass[12pt]{iopart}
\usepackage{graphicx}
\usepackage{epstopdf}

\newcommand{\binomial}[2]
{
  \left( \begin{array}{c} #1 \\ #2 \end{array} \right )
}
\newcommand{\ninej}[9]
{
  \left\{ \begin{array}{ccc} #1 & #2 & #3 \\
    #4 & #5 & #6 \\ #7 & #8 & #9 \end{array} \right\}
}
\newcommand{\sixj}[6]
{
  \left\{ \begin{array}{ccc} #1 & #2 & #3 \\
    #4 & #5 & #6 \end{array} \right\}
}

\begin{document}

\title{Four-body long-range interactions between ultracold weakly-bound diatomic molecules}

\author{M. Lepers, G. Qu{\'e}m{\'e}ner, E. Luc-Koenig and O. Dulieu}
\address{Laboratoire Aim{\'e} Cotton, CNRS -- Universit{\'e} Paris-Sud -- ENS Cachan, 91405 Orsay, France}
\ead{maxence.lepers@u-psud.fr}

\begin{abstract}
Using the multipolar expansion of electrostatic and magnetostatic potential energies, we characterize the long-range interactions between two weakly-bound diatomic molecules, taking as an example the paramagnetic Er$_2$ Feshbach molecules which were produced recently. Since inside each molecule,  individual atoms conserve their identity, the intermolecular potential energy can be expanded as the sum of pairwise atomic potential energies. In the case of Er$_2$ Feshbach molecules, we show that the interaction between atomic magnetic dipoles gives rise to the usual $R^{-3}$ term of the multipolar expansion, with $R$ the intermolecular distance, but also to additional terms scaling as $R^{-5}$, $R^{-7}$, and so on. Those terms are due to the interaction between effective molecular multipole moments, and are strongly anisotropic with respect to the orientation of the molecules. Similarly the atomic pairwise van der Waals interaction results in $R^{-6}$, $R^{-8}$, ... terms in the intermolecular potential energy. By calculating the reduced electric-quadrupole moment of erbium ground level $\langle J=6||\hat{Q}_2||J=6\rangle = -1.305$ a.u., we also demonstrate that the electric-quadrupole interaction energy is negligible with respect to the magnetic-dipole and van der Waals interaction energies. The general formalism presented in this article can be applied to calculate the long-range potential energy between arbitrary charge distributions composed of almost free subsystems.
\end{abstract}


\section{Introduction}

For a long time few-body effects have been attracting a lot of interest, especially in nuclear physics \cite{hammer2013}, resulting in some striking theoretical predictions like the Efimov effect \cite{efimov1970}. More recently the tremendous progress for controlling interactions between ultracold atoms, has allowed to experimentally confirm those predictions \cite{blume2012}. Indeed the signature of triatomic Efimov states was identified in an ultracold Bose gas of cesium, where the formation of two-body bound states was either forbidden \cite{kraemer2006} or permitted \cite{knoop2009}. An essential feature of Efimov bound states is their universality in the following sense \cite{braaten2006, gross2009, zaccanti2009}: they are characterized by two parameters, the two-body scattering length and the three-body parameter \cite{berninger2011, naidon2014}, which accounts for all the details of atomic interactions. Using ultracold atomic and molecular gases, many extensions of Efimov's original prediction \cite{efimov1970} were then explored or proposed, like four-body \cite{stecher2009, pollack2009, ferlaino2009}, or more-body bound states \cite{zenesini2013}, the impact of Fermi statistics \cite{lompe2010, serwane2011} or distinguishable particles \cite{nakajima2011}, deviation from universality \cite{lompe2010, nakajima2010, zenesini2014}, or bound states of dipolar particles \cite{wang2011}. Beside Efimov effect, the theoretical modeling of collisions involving weakly-bound dimers revealed the enhanced stability against collisions of molecules composed of identical fermions with respect to those composed of bosons \cite{petrov2004, petrov2005}. Besides, few-body collisions in the presence of dipole-dipole interactions were also explored \cite{ticknor2010}.

In this respect the production of ultracold gases of lanthanide atoms is extremely promising \cite{mcclelland2006, lu2010, sukachev2010, miao2013}. Firstly their strong magnetic dipole moment creates anisotropic and long-range dipolar interactions that, unlike electric-dipolar interactions, do not need to be induced by an external field. Secondly, erbium and dysprosium possess stable bosonic and fermionic isotopes, that were driven to quantum denegeracy \cite{lu2011b, aikawa2012, lu2012, aikawa2014, tang2015}. Despite the absence of hyperfine structure, bosonic erbium and dysprosium present dense spectra of Feshbach resonances \cite{maier2015}, which were recently used to form very weakly-bound Er$_2$ molecules, \textit{i.e.} so-called Feshbach molecules \cite{frisch2015}, through magneto-association technique. Such Er$_2$ molecules look like excellent candidates to study few-body physics in dipolar systems. 

In many investigations of few-body physics in ultracold gases, atomic interactions are described with model potentials, \textit{e.g.}~contact potentials, which is justified for atoms interacting through van der Waals forces. However if the atoms carry a magnetic dipole moment, the resulting long-range and anisotropic dipolar interaction requires a cautious modeling which takes into accounts the internal structure of the atoms. In this article, using the multipolar expansion in inverse powers of the intermolecular distance $R$ \cite{stone1996, kaplan2006}, we characterize the long-range interactions between two weakly-bound diatomic molecules. We focus on the regime where the two molecules are approaching each other, that is to say when the intermolecular distance is larger than the mean interatomic distance inside each molecule. Assuming that  individual atoms keep their own identity within each molecule, we expand the intermolecular potential energy as the sum of pairwise atomic interaction energies.
Taking the example of two Er$_2$ Feshbach  molecules, we show that, when expressed in the coordinate system of the molecule-molecule complex, the total dipole-dipole and van der Waals interactions between all atom pairs are both expressed as a sum of terms proportional to inverse powers of $R$, and involving effective molecular multipole moments. These terms, which are absent in the usual multipolar expansion, are strongly anisotropic with respect to the orientation of the two molecules. In addition, by calculating the electric-quadrupole moment of erbium ground level we show that the total quadrupolar interaction, which would in principle give rise to another series of $R^{-n}$ terms, is actually much smaller than the dipolar and van der Waals interactions. We will calculate adiabatic potential-energy curves that could be used in a future work to study Er$_2$-Er$_2$ collisions in the ultracold regime.


The article is outlined as follows. In Section \ref{sec:gene} we present the general formalism giving the first-order and second-order energy corrections between arbitrary weakly-bound charge distributions. This formalism appears as a generalization of the usual multipolar expansion. Then in Section \ref{sec:expl-FM} we consider the example of Er$_2$ Feshbach molecules, focusing on their magnetic-dipole and van der Waals interactions, and discussing also their electric-quadrupole interactions. With arguments based on the characteristic lengths associated with the multipolar expansion, we show that the anisotropic terms due to effective molecular multipole moments, are likely to play an important role in the Er$_2$-Er$_2$ collisional dynamics at ultralow energies. Section \ref{sec:conclu} contains concluding remarks.

\section{Potential energy between distant weakly-bound molecules \label{sec:gene}}

\begin{figure}
\begin{center}
\includegraphics[width=16cm]{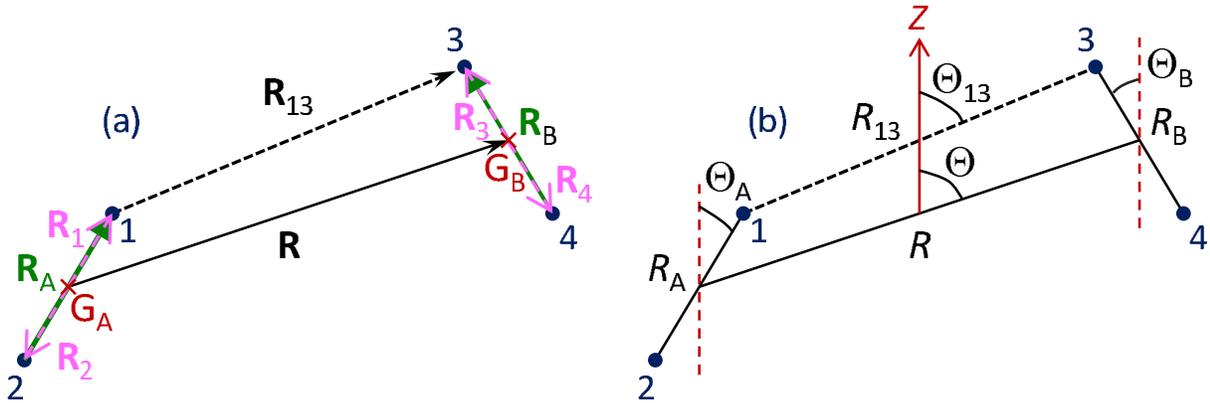}
\caption{\label{fig:CS} (Color online) (a) 
Vectors, (b) distances and angles describing the position of the weakly-bound diatomic molecules A (composed of atoms 1 and 2) and B (composed of atoms 3 and 4) in the space-fixed frame $XYZ$ with quantization axis $Z$. The relevant coordinates describing a pair of atoms from A and B is drawn, with the example of atoms 1 and 3. For sake of clarity identical atoms are considered, and the azimuthal angles are not represented.
}
\end{center}
\end{figure}

We consider two molecules denoted A and B. Molecule A is composed of atoms 1 and 2, and molecule B of atoms 3 and 4. The orientation of the interatomic axes of A and B in the space-fixed (SF) frame $XYZ$, $Z$ being the quantization axis, are characterized by the vectors $\mathbf{R}_A\equiv \mathbf{R}_1-\mathbf{R}_2$ and $\mathbf{R}_B\equiv \mathbf{R}_3-\mathbf{R}_4$, with spherical coordinates 
$(R_A, \Theta_A, \Phi_A)$ and $(R_B, \Theta_B, \Phi_B)$ respectively. The orientation of the intermolecular axis, which joins the centers of mass of A and B is given by the vector $\mathbf{R}\equiv(R, \Theta, \Phi)$ (see Fig.~\ref{fig:CS}).

\subsection{First-order correction \label{sub:1st-ord}}

When the distance $R$ goes to infinity, molecules A and B are independent from each other. Their quantum states, called $|A\rangle$ and $|B\rangle$, are characterized by the zeroth-order energies $E_A$ and $E_B$ respectively. As $R$ decreases, A and B start to interact through electrostatic and/or magnetostatic forces. We assume that within the weakly-bound molecules, each individual atom keeps its identity -- namely that the exchange is neglected -- so that atoms interact through electrostatic and/or magnetostatic forces as well. The total potential energy of the complex $V_\mathrm{tot}$ can therefore be expanded as the sum of pairwise atomic energies, $V_\mathrm{tot} = V_{12}+V_{34} + \sum_{i=1}^{2}\sum_{j=3}^{4} V_{ij}$. In this work, assuming that the first two terms $V_{12}$ and $V_{34}$ are part of the unperturbed energies $E_A$ and $E_B$, we focus on the intermolecular potential energy
\begin{equation}
  V(\mathbf{R}_A, \mathbf{R}_B, \mathbf{R})
   = \sum_{i=1}^{2}\sum_{j=3}^{4} V_{ij}(\mathbf{R}_{ij}) ,
  \label{eq:sum-ij}
\end{equation}
where $\mathbf{R}_{ij}$ is the vector pointing from atoms $i\in A$ to $j\in B$.

In Eq.~(\ref{eq:sum-ij}) the interatomic energy $V_{ij}$ is given by the usual multipolar expansion in the SF frame
\begin{equation}
  V_{ij}(\mathbf{R}_{ij}) = \sum_{\ell_{ij}=0}^{+\infty}
    \sum_{m_{ij}=-\ell_{ij}}^{+\ell_{ij}} F_{\ell_{ij}m_{ij}}
    G_{\ell_{ij}m_{ij}}^*(\mathbf{R}_{ij}),
  \label{eq:v-ij}
\end{equation}
where the factor $F_{\ell_{ij}m_{ij}}$ only depends on the atomic multipole moments,
\begin{eqnarray}
  F_{\ell_{ij}m_{ij}} & = & F_0 \sum_{\ell_i,\ell_j=0}^{\ell_{ij}}
   \delta_{\ell_i+\ell_j,\ell_{ij}}\, (-1)^{\ell_j}
   \binomial{2\ell_{ij}}{2\ell_i} ^{1/2}
   \nonumber \\
  & & \times \sum_{m_i=-\ell_i}^{+\ell_i} \sum_{m_j=-\ell_j}^{+\ell_j}
   \mathcal{C}_{\ell_i m_i \ell_j m_j}
              ^{\ell_{ij}m_{ij}}
   Q_{\ell_i m_i} Q_{\ell_j m_j}
  \label{eq:f-ij}
\end{eqnarray}
with $F_0=1/4\pi\epsilon_0$ ($\mu_0/4\pi$) for electrostatic (magnetostatic) interactions, $\epsilon_0$ and $\mu_0$ the permitivity and permeability of the vacuum, $(:)$ a binomial coefficient, $\mathcal{C}_{\ell_i m_i \ell_j m_j}^{\ell_{ij}m_{ij}} =\langle\ell_i m_i \ell_j m_j|\ell_i\ell_j\ell_{ij}m_{ij}\rangle$ a Clebsch-Gordan coefficient, and $Q_{\ell_i m_i}$ ($Q_{\ell_j m_j}$) the multipole moment of atom $i$ ($j$), expressed in a coordinate system (CS) of the SF frame and centered on atom $i$ ($j$). The multipolar expansion can be applied if all interatomic distances are larger than the so-called LeRoy radius \cite{leroy1974}, so that their electronic clouds do not overlap.

\begin{table}
  \caption{Coordinates of the vectors $\mathbf{R}_1$, $\mathbf{R}_2$, $\mathbf{R}_3$ and $\mathbf{R}_4$, as well as the geometric factors $\eta_1$, $\eta_2$, $\eta_3$ and $\eta_4$ appearing in the intermolecular potential energy (\ref{eq:sum-ij-2}), given as functions of the coordinates of $\mathbf{R}_A$ and $\mathbf{R}_B$, and of the atomic masses $\mathcal{M}_1$, $\mathcal{M}_2$, $\mathcal{M}_3$ and $\mathcal{M}_4$.
  \label{tab:comp-r-i-j}
  }
  \begin{center}
  \begin{tabular}{c|llll}
  \hline
   atom & $R_i$, $R_j$ & $\Theta_i$, $\Theta_j$
    & $\Phi_i$, $\Phi_j$ & $\eta_i$, $\eta_j$ \\
  \hline
     1  & $R_1=\frac{\mathcal{M}_2}
               {\mathcal{M}_1+\mathcal{M}_2}R_A$ 
    & $\Theta_1=\Theta_A$ & $\Phi_1=\Phi_A$ 
    & $\eta_1=+\frac{\mathcal{M}_2}
               {\mathcal{M}_1+\mathcal{M}_2}$ \\
     2  & $R_2=\frac{\mathcal{M}_1}
               {\mathcal{M}_1+\mathcal{M}_2}R_A$ 
    & $\Theta_2=\pi-\Theta_A$ & $\Phi_2=\Phi_A+\pi$
    & $\eta_2=-\frac{\mathcal{M}_1}
               {\mathcal{M}_1+\mathcal{M}_2}$ \\
     3  & $R_3=\frac{\mathcal{M}_4}
               {\mathcal{M}_3+\mathcal{M}_4}R_B$ 
    & $\Theta_3=\Theta_B$ & $\Phi_3=\Phi_B$
    & $\eta_3=+\frac{\mathcal{M}_4}
              {\mathcal{M}_3+\mathcal{M}_4}$ \\
     4  & $R_4=\frac{\mathcal{M}_3}
               {\mathcal{M}_3+\mathcal{M}_4}R_B$ 
    & $\Theta_4=\pi-\Theta_B$ & $\Phi_4=\Phi_B+\pi$
    & $\eta_4=-\frac{\mathcal{M}_3}
               {\mathcal{M}_3+\mathcal{M}_4}$ \\
  \hline
  \end{tabular}
  \end{center}
\end{table}

In Eq.~(\ref{eq:v-ij}), $G_{\ell_{ij}m_{ij}}$ is a purely geometric factor involving the Racah spherical harmonics $C_{\ell_{ij}m_{ij}}(\Theta_{ij},\Phi_{ij})$.
\begin{equation}
  G_{\ell_{ij}m_{ij}}(\mathbf{R}_{ij}) 
   = \frac{C_{\ell_{ij}m_{ij}}(\Theta_{ij},\Phi_{ij})}
          {R_{ij}^{1+\ell_{ij}}}.
  \label{eq:g-ij}
\end{equation}
It depends on the relative coordinates of atoms $i$ and $j$, which are not compatible to the CS defined in Figure \ref{fig:CS}. In order to express $G_{\ell_{ij}m_{ij}}$ in this CS we use the following relation
\begin{equation}
  \mathbf{R}_{ij} = \mathbf{R} - (\mathbf{R}_i-\mathbf{R}_j)
             \equiv \mathbf{R} -  \mathbf{U}_{ji} \,,
\end{equation}
where $\mathbf{R}_i$ ($\mathbf{R}_j$) are the vectors joining the center of mass of molecule A (B) to the atom $i$ ($j$). Those vectors are colinear to $\mathbf{R}_A$ ($\mathbf{R}_B$) and their coordinates are given in Table \ref{tab:comp-r-i-j}. The coordinates of $\mathbf{U}_{ji}$ are called $(U_{ji}, \Xi_{ji}, \Psi_{ji})$.

To transform Eq.~(\ref{eq:g-ij}), we expand the spherical harmonics for the vector $\mathbf{r} = \mathbf{r}' - \mathbf{r}'' = \mathbf{R}_{ij}$ as function of those for the vectors $\mathbf{r}' = \mathbf{R}$ and $\mathbf{r}'' = \mathbf{U}_{ji}$. Setting $\mathbf{r}\equiv(r,\theta,\phi)$ (and similarly for $\mathbf{r}'$ and $\mathbf{r}''$), we apply (see Ref.~\cite{varshalovich1988}, Ch.~5, Eq.~(36), p.~167 \footnote{In \cite{varshalovich1988} the relation is given in terms of normalized spherical harmonics $Y_{kq}$; we transform it using $Y_{kq}(\theta,\phi) = \sqrt{(2k+1)/4\pi}\times C_{kq}(\theta,\phi)$})
\begin{eqnarray}
  \frac{C_{kq}(\theta,\phi)}{r^{1+k}} & = &
    \sum_{k',k''=0}^{+\infty} \delta_{k'-k'',k} \, (-1)^{k'} 
    \left ( \frac{(2k'+1)!}{(2k'')!(2k+1)!}\right )^{1/2}
    \frac{(r'')^{k''}}{(r')^{k'+1}}
  \nonumber \\
    & & \times \sum_{q'=-k'}^{+k'} \sum_{q''=-k''}^{+k''}
    \mathcal{C}_{k''q''k'q'}^{kq} 
    C_{k''q''}(\theta'',\phi'')
    C_{k'q'}(\theta',\phi')
  \label{eq:Ckq-1}
\end{eqnarray}
which is valid for $r''<r'$. We obtain for Eq.~(\ref{eq:g-ij})
\begin{eqnarray}
  G_{\ell_{ij}m_{ij}}(\mathbf{R}_{ij}) & = & 
    \sum_{\lambda=\ell_{ij}}^{+\infty} (-1)^{\lambda} 
    \left ( \frac{(2\lambda+1)!}
                 {(2\lambda-2\ell_{ij})!(2\ell_{ij}+1)!}
    \right )^{1/2}
    \frac{U_{ji}^{\lambda-\ell_{ij}}}{R^{1+\lambda}}
    \nonumber \\
     & & \times \sum_{\mu=-\lambda}^{+\lambda}
    \mathcal{C}^{\ell_{ij}m_{ij}}
     _{\lambda-\ell_{ij},\mu-m_{ij},\lambda\mu} 
    C_{\lambda-\ell_{ij},\mu-m_{ij}}
      (\Xi_{ji},\Psi_{ji})
    C_{\lambda\mu}(\Theta,\Phi)\,.
  \label{eq:g-ij-2}
\end{eqnarray}
which is then valid for $U_{ji}<R$. Since $0\le U_{ji}\le R_i+R_j$, Eq.~(\ref{eq:g-ij-2}) is applicable provided that
\begin{equation}
  R_i + R_j < R \,.
  \label{eq:valid-ij}
\end{equation}
We will come back to this criterion later on.

Equation (\ref{eq:g-ij-2}) represents the first main result of our approach: the pairwise atomic potential energy has been transformed from a sum of terms in the relative CS of the interacting atoms (see Eq.~(\ref{eq:g-ij})), into a sum of terms proportional to inverse powers of the intermolecular distance $R$ (see Eq.~(\ref{eq:g-ij-2})). The price to pay is the emergence, for each couple of atomic multipole moments $(\ell_i,\ell_j)$, of an infinite sum of terms scaling as $R^{-1-\lambda}$. Each term is anisotropic due to the Racah spherical harmonics $C_{\lambda\mu}(\Theta,\Phi)$.
To express the energy as a function of $\mathbf{R}_i$ and $\mathbf{R}_j$, we apply a second transformation (see Ref.~\cite{varshalovich1988}, Ch.~5, Eq.~(35)) to $\mathbf{r}=\mathbf{U}_{ji}$, $\mathbf{r'} = \mathbf{R}_i$ and $\mathbf{r''} = \mathbf{R}_j$
\begin{eqnarray}
  r^k C_{kq}(\theta,\phi) & = & \sum_{k',k''=0}^{k}
    \delta_{k'+k'',k} \,(-1)^{k''}
    \left ( \frac{(2k)!}{(2k')!(2k'')!} \right ) ^{1/2}
    (r')^{k'} (r'')^{k''}
  \nonumber \\
    & & \times \sum_{q'=-k'}^{+k'} \sum_{q''=-k''}^{+k''}
    \mathcal{C}_{k'q'k''q''}^{kq} 
    C_{k'q'}(\theta',\phi')
    C_{k''q''}(\theta'',\phi''),
\end{eqnarray}
which is valid for all $r'$ and $r''$. This gives the final expression for $G_{\ell_{ij}m_{ij}}$ 
\begin{eqnarray}
  G_{\ell_{ij}m_{ij}}(\mathbf{R}_{ij}) & = & 
    \sum_{\lambda=\ell_{ij}}^{+\infty}
    \sum_{\lambda_i,\lambda_j=0}^{\lambda-\ell_{ij}}
    \delta_{\ell_{ij}+\lambda_i+\lambda_j,\lambda}
    \, (-1)^{\lambda+\lambda_j}
    \left [ \binomial{2\lambda+1}{2\ell_{ij}+1}
            \binomial{2\lambda_i+2\lambda_j}{2\lambda_i}
    \right ] ^{1/2}
  \nonumber \\
    & & \times \frac{R_i^{\lambda_i}R_j^{\lambda_j}}
                    {R^{1+\lambda}}
    \sum_{\mu=-\lambda}^{+\lambda}
    \mathcal{C}_{\lambda-\ell_{ij},\mu-m_{ij},\lambda\mu}^{\ell_{ij}m_{ij}}
    C_{\lambda\mu}(\Theta,\Phi)
  \nonumber \\
    & & \times \sum_{\mu_i=-\lambda_i}^{+\lambda_i}
    \sum_{\mu_j=-\lambda_j}^{+\lambda_j}
    \mathcal{C}_{\lambda_i\mu_i\lambda_j\mu_j}
     ^{\lambda-\ell_{ij},\mu-m_{ij}}
    C_{\lambda_i\mu_i}(\Theta_i,\Phi_i)
    C_{\lambda_j\mu_j}(\Theta_j,\Phi_j)
  \label{eq:g-ij-3}
\end{eqnarray}
In addition to the sum over $\lambda$, we get two sums over $\lambda_i$ and $\lambda_j$ which, due to their $R_i^{\lambda_i} C_{\lambda_i\mu_i}(\Theta_i,\Phi_i)$ dependence (and similarly for $j$), can be viewed as effective multipole moments describing the position of atoms $i$ and $j$ in the CS associated with A and B respectively. It is worthwile mentioning that for $R_i,R_j\ll R$, namely when the size of the molecules is negligible with respect to their mutual distance, the sums reduce to $\lambda_i=\lambda_j=0$ and to $\lambda=\ell_{ij}$, and so we recover the usual multipolar expansion.

Finally, when adding up all pairwise atomic contributions, we can replace $\lambda_i$ by $\lambda_A$ and $\lambda_j$ and $\lambda_B$, as the position of atoms 1 and 3 in the CS of A and B gives the orientation of the interatomic axes of A and B respectively. For atoms 2 and 4, the orientation is opposite to that of the corresponding interatomic axes, and so we use $C_{kq}(\pi-\theta,\phi+\pi)=(-1)^k C_{kq}(\theta,\phi)$. Moreover replacing $\ell_{ij}$ by $\ell$ for simplicity, and using $C_{kq}^*(\theta,\phi)=(-1)^q C_{k,-q}(\theta,\phi)$ and the particular expression of $\mathcal{C}_{k'q'k''q''}^{kq}$ for $k=k'\pm k''$ \cite{varshalovich1988}, we obtain for the intermolecular potential energy of Eq.~(\ref{eq:sum-ij})
\begin{eqnarray}
  V(\mathbf{R}_A, \mathbf{R}_B, \mathbf{R}) & = & F_0
    \sum_{\ell=0}^{+\infty} \sum_{\lambda=\ell}^{+\infty}
    \sum_{\lambda_A,\lambda_B=0}^{\lambda-\ell}
    \delta_{\lambda_A+\lambda_B+\ell,\lambda} 
    \, (-1)^{\lambda_B}
    \frac{R_A^{\lambda_A}R_B^{\lambda_B}}{R^{1+\lambda}}
  \nonumber \\
    & \times & \sum_{\mu=-\lambda}^{+\lambda}
    \sqrt{(\lambda+\mu)!(\lambda-\mu)!}
    \times C^*_{\lambda\mu}(\Theta,\Phi)
  \nonumber \\
    & \times & \sum_{\mu_A=-\lambda_A}^{+\lambda_A}
               \sum_{\mu_B=-\lambda_B}^{+\lambda_B}
    \frac{C_{\lambda_A\mu_A}(\Theta_A,\Phi_A)
          C_{\lambda_B\mu_B}(\Theta_B,\Phi_B)}
         {\sqrt{(\lambda_A+\mu_A)!(\lambda_A-\mu_A)!
                (\lambda_B+\mu_B)!(\lambda_B-\mu_B)!}}
  \nonumber \\
    & \times & \sum_{i=1}^{2}\sum_{j=3}^{4}
    \eta_i^{\lambda_A} \eta_j^{\lambda_B}
    \sum_{\ell_i,\ell_j=0}^{\ell}
    \delta_{\ell_i+\ell_j,\ell} \,(-1)^{\ell_i}
    \nonumber \\
    & \times & \sum_{m_i=-\ell_i}^{+\ell_i} 
               \sum_{m_j=-\ell_j}^{+\ell_j}
    \frac{\delta_{\mu_A+\mu_B+m_i+m_j,\mu}
          Q_{\ell_i m_i} Q_{\ell_j m_j}}
         {\sqrt{(\ell_i+m_i)!(\ell_i-m_i)!
                (\ell_j+m_j)!(\ell_j-m_j)!}}
  \label{eq:sum-ij-2}
\end{eqnarray}
where the geometric factors $\eta_i$ and $\eta_j$ are given in Table \ref{tab:comp-r-i-j}. The Kronecker symbol in the last line of Eq.~(\ref{eq:sum-ij-2}) is obtained by combining the Clebsch-Gordan coefficients of Eqs.~(\ref{eq:f-ij}) and (\ref{eq:g-ij-3}). Since the condition $\mu_A+\mu_B+m_i+m_j = \mu$ must be satisfied for all pairs $(i,j)$, it imposes that $m_1=m_2$ and $m_3=m_4$. 

As already mentioned, Eq.~(\ref{eq:sum-ij-2}) can be viewed as the interaction between effective multipole moments of ranks $\lambda_A$ and $\lambda_B$, describing the orientation of the interatomic axes of molecules A and B. Equation (\ref{eq:sum-ij-2}) comes out as the usual sum of terms proportional to inverse powers of the intermolecular distance $R^{1+\lambda}$, and to angular factors $C_{\lambda\mu}(\Theta,\Phi)$, which account for the anisotropy of long-range interactions. Since $-1\le\eta_i, \eta_j\le 1$, Eq.~(\ref{eq:valid-ij}) can be rewritten as a condition of validity for Eq.~(\ref{eq:sum-ij-2})
\begin{equation}
  \max(|\eta_1|,|\eta_2|)R_A + \max(|\eta_3|,|\eta_4|)R_B < R.
  \label{eq:valid-AB}
\end{equation}
In the homonuclear case, \textit{e.g.}~$\mathcal{M}_1 = \mathcal{M}_2$, $\max(|\eta_1|,|\eta_2|)=1/2$. On the contrary if $\mathcal{M}_1 \gg \mathcal{M}_2$ or $\mathcal{M}_1 \ll \mathcal{M}_2$, then $\max(|\eta_1|,|\eta_2|)=1$.

\subsection{Second-order correction \label{sub:2nd-ord}}

Calculating the second-order correction to the intermolecular potential-energy (\ref{eq:sum-ij}) is equivalent to calculating the matrix elements between unperturbed states of the second-order operator
\begin{equation}
  W(\mathbf{R}_A, \mathbf{R}_B, \mathbf{R}) 
    = -\sum_{(A'',B'')\neq(A,B)}
    \frac{V|A''B''\rangle\langle A''B''|V}
         {E_{A''}-E_A+E_{B''}-E_B}
  \label{eq:2ord-oper}
\end{equation}
where $|A''\rangle$ and $|B''\rangle$ formally denote the states of molecules $A$ and $B$ which are coupled to $|A\rangle$ and $|B\rangle$ by $V$. The sum is performed for all possible pairs of states $(|A''\rangle,|B''\rangle)$ excluding the case where both $|A''\rangle = |A\rangle$ and $|B''\rangle = |B\rangle$. Because the unperturbed states $|A\rangle$ and $|B\rangle$ correspond to weakly-bound molecules, we can suppose that the states $|A''B''\rangle$ such that $\langle AB|\hat{V}|A''B''\rangle\neq 0$ also correspond to weakly-bound molecules, but near different atomic dissociation limits. Therefore in Eq.~(\ref{eq:2ord-oper}) the energy differences $E_{A''}-E_A$ and $E_{B''}-E_B$ can be replaced by the energy differences between the corresponding atomic dissociation limits. We can replace the sum over the molecular states $|A''\rangle$ and $|B''\rangle$ by a sum over states of the separated atoms $|1''\rangle$, $|2''\rangle$, $|3''\rangle$ and $|4''\rangle$.
This assumption implies that the geometric factors $G_{\ell_{ij}m_{ij}}$ given by Eq.~(\ref{eq:g-ij-3}) will be taken out of the sum over atomic states, which gives for Eq.~(\ref{eq:2ord-oper})
\begin{eqnarray}
  W(\mathbf{R}_A, \mathbf{R}_B, \mathbf{R}) & = &
    -\sum_{i,i'=1}^{2} \sum_{j,j'=3}^{4}
    \sum_{\ell_{i j },\ell_{i'j'}=0}^{+\infty}
    \sum_{m_{i j }=-\ell_{i j }}^{+\ell_{ij}}
    \sum_{m_{i'j'}=-\ell_{i'j'}}^{+\ell_{i'j'}}
    G_{\ell_{i j }m_{i j }}^*(\mathbf{R}_{i j })    
    G_{\ell_{i'j'}m_{i'j'}}^*(\mathbf{R}_{i'j'})    
  \nonumber \\
  & \times & \sum_{1'',2'',3'',4''}
    \frac{F_{\ell_{ij}m_{ij}}|1'',2'',3'',4''\rangle
          \langle 1'',2'',3'',4''|F_{\ell_{i'j'}m_{i'j'}}}
         {\Delta_{1''}+\Delta_{2''}+\Delta_{3''}+\Delta_{4''}}
  \,, \label{eq:2ord-oper-2}
\end{eqnarray}
where $\Delta_{k''}=E_{k''}-E_k$ is the excitation energy of atom $k$ ($k=1$ to 4). The sum over atomic states excludes the case $[k''\rangle = |k\rangle$ for all the atoms at the same time.

Applying Eq.~(\ref{eq:2ord-oper-2}) with the particular form of $F_{\ell_{ij}m_{ij}}$ and $G_{\ell_{ij}m_{ij}}$ given by Eqs.~(\ref{eq:f-ij}) and (\ref{eq:g-ij-3}) would be inconvenient, as it would not yield terms with irreducible tensors, and so would prevent from using the Wigner-Eckart theorem for the matrix elements of $W$. Insted we introduce coupled multipole moments of rank $k_A$ associated with atoms $i$ and $i'$ (see e.g.~\cite{spelsberg1993, bussery-honvault2008, lepers2012})
\begin{equation}
  \mathcal{Q}_{(\ell_i,\ell_{i'})k_A q_A}'' = 
    \sum_{m_{i }=-\ell_{i }}^{+\ell_{i }}
    \sum_{m_{i'}=-\ell_{i'}}^{+\ell_{i'}}
      \mathcal{C}_{\ell_{i}m_{i}\ell_{i'}m_{i'}}^{k_A q_A}
      Q_{\ell_{i}m_{i}} |1''2''\rangle
  \langle 1''2''| Q_{\ell_{i'}m_{i'}} 
  \label{eq:cpl-mult}
\end{equation}
and the same for $j$, $j'$ and B. Doing similar transformations for Racah spherical harmonics (see \ref{sec:app} for details), we obtain the final expression for $W$
\begin{eqnarray}
  W(\mathbf{R}_A, \mathbf{R}_B, \mathbf{R}) & = & -F_0^2
    \sum_{\ell \lambda \lambda_A \lambda_B }
    \sum_{\ell'\lambda'\lambda_A'\lambda_B'}
    \delta_{\lambda_A'+\lambda_B'+\ell',\lambda'}
    \delta_{\lambda_A +\lambda_B +\ell ,\lambda }\,
    \frac{R_A^{\lambda_A+\lambda_A'}
          R_B^{\lambda_B+\lambda_B'}}
         {R^{2+\lambda+\lambda'}}
  \nonumber \\
    & \times & \left [
                 \binomial{2\lambda+1}{2\ell+1}
                 \binomial{2\lambda_A+2\lambda_B}{2\lambda_A}
                 \binomial{2\lambda'+1}{2\ell'+1}
                 \binomial{2\lambda_A'+2\lambda_B'}{2\lambda_A'}
    \right ] ^{1/2}
  \nonumber \\
    & \times & \sum_{k\kappa\pi\kappa_A\kappa_B}
    (-1)^{\kappa+\kappa_B} \times
    [ (\lambda_A+\lambda_B)(\lambda_A'+\lambda_B')
      \kappa_A\kappa_B\kappa\pi ]^{1/2}
  \nonumber \\
    & \times &
    \ninej{\lambda_A }{\lambda_B }{\lambda_A +\lambda_B }
          {\lambda_A'}{\lambda_B'}{\lambda_A'+\lambda_B'}
          { \kappa_A }{ \kappa_B }{\pi}
    \ninej{\lambda_A +\lambda_B }{\lambda }{\ell }
          {\lambda_A'+\lambda_B'}{\lambda'}{\ell'}
          {\pi}                  { \kappa }{k}
    \mathcal{C}_{\lambda_A0\lambda_A'0}^{\kappa_A0}
    \mathcal{C}_{\lambda_B0\lambda_B'0}^{\kappa_B0}
    \mathcal{C}_{\lambda  0\lambda'  0}^{\kappa  0}
  \nonumber \\
    & \times & \sum_{q\varrho\sigma\varrho_A \varrho_B}
    \mathcal{C}_{\kappa_A\varrho_A\kappa_B\varrho_B}^{\pi\sigma}
    \mathcal{C}_{\pi\sigma\kappa\varrho}^{kq}
    C_{\kappa  \varrho  }^*(\Theta  ,\Phi  )
    C_{\kappa_A\varrho_A}^*(\Theta_A,\Phi_A)
    C_{\kappa_B\varrho_B}^*(\Theta_B,\Phi_B)
  \nonumber \\
  & \times & \sum_{ii'jj'}
    \eta_i^{\lambda_A} \eta_{i'}^{\lambda_A'}
    \eta_j^{\lambda_B} \eta_{j'}^{\lambda_B'}
    \sum_{\ell_{i}\ell_{i'}\ell_{j}\ell_{j'}}
    \delta_{\ell_{i }+\ell_{j },\ell }
    \delta_{\ell_{i'}+\ell_{j'},\ell'}
    (-1)^{\ell_j+\ell_{j'}}
  \nonumber \\
  & \times & \left[
               \binomial{2\ell }{2\ell_{i }}
               \binomial{2\ell'}{2\ell_{i'}} \right] ^{1/2}
    [\ell\ell'] \sum_{k_A k_B} [k_A k_B]^{1/2}
    \ninej{\ell_{i }}{\ell_{j }}{\ell }
          {\ell_{i'}}{\ell_{j'}}{\ell'}
          {   k_A  }{    k_B  }{k     }
  \nonumber \\
  & \times & \sum_{q_A q_B} \mathcal{C}_{k_A q_A k_B q_B}^{kq}
    \sum'_{1'',2'',3'',4''}
    \frac{\mathcal{Q}_{(\ell_i,\ell_{i'})k_A q_A}''
          \mathcal{Q}_{(\ell_j,\ell_{j'})k_B q_B}''}
         {\Delta_{1''}+\Delta_{2''}+\Delta_{3''}+\Delta_{4''}}\,,
  \label{eq:2ord-oper-3}
\end{eqnarray}
This equation contains several sums: over all the quantum states $|1''\rangle$, $|2''\rangle$, $|3''\rangle$ and $|4''\rangle$ of the four atoms, on the atoms themselves ($i,i'=1,2$ and $j,j'=3,4$), and over tensor-operator ranks and components. In this respect the Latin letters correspond to the atomic multipole moments, and the Greek ones to the effective molecular multipole moments. The letters $\ell$ and $\lambda$ characterize uncoupled tensor operators, whereas $k$, $\kappa$ and $\pi$ characterize coupled ones (see below), and $q$, $\varrho$ and $\sigma$ their components. The unprimed (primed) uncoupled tensor ranks come from the first (second) call of the multipolar operator $V$ in the second-order correction (see Eq.~(\ref{eq:2ord-oper})).

\begin{table}[h!]
  \caption{Mathematical and physical definitions of the ranks of the coupled tensors appearing in Eq.~(\ref{eq:2ord-oper-3}). They are constructed by vector addition in the sense of angular-momentum theory.
  \label{tab:cpl-tens}
  }
  \begin{center}
  \begin{tabular}{ll}
  \hline
   tensor rank & physical quantity \\
  \hline
   $\vec{k}_A = \vec{\ell}_i + \vec{\ell}_{i'}$
    & coupled multipole moment of atoms $(i,i')$ \\
   $\vec{k}_B = \vec{\ell}_j + \vec{\ell}_{j'}$
    & coupled multipole moment of atoms $(j,j')$ \\
   $\vec{\kappa}_A = \vec{\lambda}_A + \vec{\lambda}_{A'}$
    & coupled effectivemultipole moment of molecule A \\
   $\vec{\kappa}_B = \vec{\lambda}_B + \vec{\lambda}_{B'}$
    & coupled effectivemultipole moment of molecule B \\
   $\vec{\kappa} = \vec{\lambda} + \vec{\lambda}'$
    & coupled tensor for the orientation of the intermolecular axis \vspace{6pt}\\
   \multicolumn{2}{l}{$\vec{\pi} = (\vec{\lambda}_A + \vec{\lambda}_{A'}) + ( \vec{\lambda}_B + \vec{\lambda}_{B'}) = \vec{\kappa}_A+\vec{\kappa}_B$} \\
   \multicolumn{2}{l}{$\vec{k} = \vec{\pi} + \vec{\kappa} = \vec{\ell} + \vec{\ell}'$} \\
  \hline
  \end{tabular}
  \end{center}
\end{table}

Equation (\ref{eq:2ord-oper-3}) shows that the second-order multipolar interaction is based on building blocks which are tensor operators, \textit{i.e.}~the atomic multipole moments $\ell_i$, $\ell_j$, $\ell_{i'}$ and $\ell_{j'}$, and the effective molecular ones $\lambda_A$, $\lambda_B$, $\lambda_{A'}$ and $\lambda_{B'}$. The $R$-dependence of the operator $W$ is obtained by adding those building blocks, namely $R^{-2-\lambda-\lambda'}$. The angular dependence of $W$ is associated with coupled tensors whose ranks are obtained by adding up the building blocks in the sense of angular momentum theory (see Table \ref{tab:cpl-tens}).

\section{Example: interactions between {Er}$_2$ Feshbach molecules \label{sec:expl-FM}}

In a recent experiment \cite{frisch2015}, an ultracold gas of bosonic $^{168}$Er atoms (with vanishing nuclear spin $I=0$) was produced in the lowest Zeeman sublevel $|J=6,M_J=-6\rangle$ of the atomic ground level [Xe]$4f^{12}6s^2$ $^3H_6$, $I=0$. A magnetic-field ramp was applied in order to transfer pairs of free atoms into a weakly-bound molecular level, thus creating a so-called Feshbach molecule. For molecule A, such a level called $|v_A\rangle$ can be expanded in the general form
\begin{equation}
  |v_A\rangle = \sum_{M_{J_1}M_{J_2}}\sum_{N_AM_{N_A}}
    \int dR_A|R_A\rangle\, \chi_{M_{J_1}M_{J_2}N_AM_{N_A}}^{v_A}(R_A) \,
    |M_{J_1}M_{J_2}N_AM_{N_A}\rangle
  \label{eq:FM}
\end{equation}
where $M_{J_1}$ and $M_{J_2}$ are the projections of total electronic angular momentum of resp.~atoms 1 and 2 on the magnetic-field axis Z, $N_A$ and $M_{N_A}$ are the angular momentum and its projection associated with the rotation of the interatomic axis of molecule A, $R_A$ is the distance between atoms 1 and 2, and $\chi_{M_{J_1}M_{J_2}N_AM_{N_A}}^{v_A}(R_A)$ is the multi-channel radial wave function describing the rovibrational motion of the molecule. The couplings between the different channels $|M_{J_1}M_{J_2}N_AM_{N_A}\rangle$ are due to the magnetic-dipole and van der Waals interactions between two erbium atoms. Since the entrance open channel corresponds to the atoms in the lowest Zeeman sublevel colliding in $s$ wave, \textit{i.e.} $M_{J_1}=M_{J_2}=-J=-6$, $N_A=M_{N_A}=0$, the allowed channels in Eq.~(\ref{eq:FM}) are such that \cite{frisch2015}: (i) $N_A$ is even and (ii) $M_{J_1}+M_{J_2}+M_{N_A} = -2J=-12$. The same selection rules apply for molecule B.

\subsection{Magnetic-dipole interaction \label{sub:mag-dip}}

Each erbium atom carries a permanent magnetic dipole moment equal to $-\mu_B g_J \vec{J}$, $\vec{J}$ being the electronic angular momentum ($J=6$), with $\mu_B$ the Bohr magneton and $g_J=1.16683\approx 7/6$ the Land{\'e} $g$-factor of erbium ground level $^3H_6$. Following Eq.~(\ref{eq:sum-ij-2}) the first-order interaction is such that $\ell_i=\ell_j=1$, $\ell=2$ and $F_0=\mu_0/4\pi$. Due to that interaction two Er$_2$ Feshbach molecules in levels $|v_A\rangle$ and $|v_B\rangle$ colliding in the partial wave $L$ and Z-projection $M_L$ can undergo elastic or inelastic scattering towards $|v'_Av'_BL'M'_L\rangle$. The matrix element of the magnetic-dipole interaction $\hat{V}_\mathrm{md}$ is then
\begin{eqnarray}
   & & \langle v'_Av'_BL'M'_L|\hat{V}_\mathrm{md}|v_Av_BLM_L\rangle
  \nonumber \\
   & = & -\frac{\mu_0}{4\pi R^3} \left(\mu_B g_J\right)^2
    J(J+1) \sum_{\lambda=2}^{+\infty}
    \sum_{\lambda_A,\lambda_B=0,2,...}^{\lambda-2}
    \delta_{\lambda_A+\lambda_B+2,\lambda} \,
    \frac{(-1)^{\lambda_B}}{(2R)^{\lambda_A+\lambda_B}}
  \nonumber \\
   & \times & \sum_{M_{J_1}M_{J_2}} \sum_{M'_{J_1}M'_{J_2}}
    \left(\frac{\delta_{M_{J_2}M'_{J_2}}
                \mathcal{C}_{JM_{J_1}1,M'_{J_1}-M_{J_1}}^{JM'_{J_1}}}
               {\sqrt{(1+M'_{J_1}-M_{J_1})!(1-M'_{J_1}+M_{J_1})!}} 
    \right.
  \nonumber \\
   & & \left.
         +\frac{\delta_{M_{J_1}M'_{J_1}}
                \mathcal{C}_{JM_{J_2}1,M'_{J_2}-M_{J_2}}^{JM'_{J_2}}}
               {\sqrt{(1+M'_{J_2}-M_{J_2})!(1-M'_{J_2}+M_{J_2})!}} 
       \right)
  \nonumber \\
   & \times & \sum_{M_{J_3}M_{J_4}} \sum_{M'_{J_3}M'_{J_4}}
    \left(\frac{\delta_{M_{J_4}M'_{J_4}}
                \mathcal{C}_{JM_{J_3}1,M'_{J_3}-M_{J_3}}^{JM'_{J_3}}}
               {\sqrt{(1+M'_{J_3}-M_{J_3})!(1-M'_{J_3}+M_{J_3})!}} 
    \right.
  \nonumber \\
   & & \left.
         +\frac{\delta_{M_{J_3}M'_{J_3}}
                \mathcal{C}_{JM_{J_4}1,M'_{J_4}-M_{J_4}}^{JM'_{J_4}}}
               {\sqrt{(1+M'_{J_4}-M_{J_4})!(1-M'_{J_4}+M_{J_4})!}} 
       \right)
  \nonumber \\
   & \times & \sum_{N_AM_{N_A}} \sum_{N'_AM'_{N_A}} 
    \int_{0}^{+\infty}dR_A \, R_A^{\lambda_A} \, 
    \chi_{M'_{J_1}M'_{J_2}N'_AM'_{N_A}}^{v'_A}(R_A)
    \,\chi_{M_{J_1}M_{J_2}N_AM_{N_A}}^{v_A}(R_A)
  \nonumber \\
   & \times & \sum_{N_BM_{N_B}} \sum_{N'_BM'_{N_B}}
    \int_{0}^{+\infty}dR_B \, R_B^{\lambda_B} \,
    \chi_{M'_{J_3}M'_{J_4}N'_BM'_{N_B}}^{v'_B}(R_B)
    \,\chi_{M_{J_3}M_{J_4}N_BM_{N_B}}^{v_B}(R_B)
  \nonumber \\
   & \times & \sqrt{(\lambda+M_L-M'_L)!(\lambda-M_L+M'_L)!}
    \sqrt{\frac{2L'+1}{2L+1}} \mathcal{C}_{L'0\lambda 0}^{L0}
    \mathcal{C}_{L'M'_L\lambda,M_L-M'_L}^{LM_L}
  \nonumber \\
   & \times & \sqrt{\frac{2N_A+1}{2N'_A+1}}
    \frac{\mathcal{C}_{N_A0\lambda_A0}^{N'_A0}
          \mathcal{C}_{N_AM_{N_A}\lambda_A,M'_{N_A}-M_{N_A}}^{N'_AM'_{N_A}}} 
          {\sqrt{(\lambda_A+M'_{N_A}-M_{N_A})!
                 (\lambda_A-M'_{N_A}+M_{N_A})!}}
  \nonumber \\
   & \times & \sqrt{\frac{2N_B+1}{2N'_B+1}}
    \frac{\mathcal{C}_{N_B0\lambda_B0}^{N'_B0}
          \mathcal{C}_{N_BM_{N_B}\lambda_B,M'_{N_B}-M_{N_B}}^{N'_BM'_{N_B}}} 
          {\sqrt{(\lambda_B+M'_{N_B}-M_{N_B})!
                 (\lambda_B-M'_{N_B}+M_{N_B})!}} \, .
  \label{eq:mag-dip}
\end{eqnarray}
Contrary to Eq.~(\ref{eq:sum-ij-2}) the terms arising from the purely atomic part of the interaction are written in the first five lines of Eq.~(\ref{eq:mag-dip}). The atomic dipole moment is expressed using the Wigner-Eckart theorem $\langle J'M'_{J_1}|\hat{Q}_{1m_1}|JM_{J_1}\rangle = -\mu_B g_J \mathcal{C}_{JM_{J_1}1m_1}^{J'M'_{J_1}} \langle J'||\hat{J}||J\rangle / \sqrt{2J+1}$ with $\langle J'||\hat{J}||J\rangle = \sqrt{J(J+1)(2J+1)}$ the reduced electronic angular momentum. For homonuclear molecules, only even values of $\lambda_A$ and $\lambda_B$ are possible, since $\eta_1=-\eta_2 = \eta_3=-\eta_4 = 1/2$. All the tensor components of Eq.~(\ref{eq:sum-ij-2}) are replaced by their only possible value, \textit{i.e.} $m_1=M'_{J_1}-M_{J_1}$, and similarly for atoms 2, 3 and 4, $\mu=M_L-M'_L$, $\mu_A=M'_{N_A}-M_{N_A}$, and similarly for B. The relationship between the tensor components established in Eq.~(\ref{eq:sum-ij-2}) imposes the following selections rules
\begin{eqnarray}
  M'_L+M'_{N_A}+M'_{N_B}+M'_{J_1}+M'_{J_3} & = &
    M_L+M_{N_A}+M_{N_B}+M_{J_1}+M_{J_3}
  \label{eq:selec-rule-1} \\
  M'_L+M'_{N_A}+M'_{N_B}+M'_{J_2}+M'_{J_3} & = &
    M_L+M_{N_A}+M_{N_B}+M_{J_2}+M_{J_3}
  \\
  M'_L+M'_{N_A}+M'_{N_B}+M'_{J_1}+M'_{J_4} & = &
    M_L+M_{N_A}+M_{N_B}+M_{J_1}+M_{J_4}
  \\
  M'_L+M'_{N_A}+M'_{N_B}+M'_{J_2}+M'_{J_4} & = &
    M_L+M_{N_A}+M_{N_B}+M_{J_2}+M_{J_4}
\end{eqnarray}
which implies
\begin{eqnarray}
  M'_{J_1}-M'_{J_2} & = & M_{J_1}-M_{J_2} \\
  M'_{J_3}-M'_{J_4} & = & M_{J_3}-M_{J_4} \,.
  \label{eq:selec-rule-2}
\end{eqnarray}
Finally in level $|v_A\rangle$ (see Eq.~(\ref{eq:FM})), the rotation of the interatomic and intermolecular axes is described by spherical harmonics $Y_{N_AM_{N_A}}(\Theta_A,\Phi_A)$, $Y_{N_BM_{N_B}}(\Theta_B,\Phi_B)$ and $Y_{LM}(\Theta,\Phi)$, and the integral of products of three spherical harmonics is used to obtain the last three lines of Eq.~(\ref{eq:mag-dip}).

Equation (\ref{eq:mag-dip}) consists in a series of inverse powers of the intermolecular distance $R$. The leading term, which scales as $R^{-3}$, appears for $\lambda_A = \lambda_B = 0$ and $\lambda=2$. It couples the channels characterized by the same $N_A$, $M_{N_A}$, $N_B$ and $M_{N_B}$, but by possibly different $M_L$ and $M_{J_{i,j}}$. Taking $M'_L = M_L$ and $M'_{J_{i,j}} = M_{J_{i,j}}$ yields the usual two-body dipole-dipole interaction (see Eq.~(\ref{eq:v-ij})) between magnetic moments $d_{v_A}$ and $d_{v_B}$ such that
\begin{equation}
  d_{v_A} = -\mu_B g_J \sum_{M_{J_1}M_{J_2}}
    w_{M_{J_1}M_{J_2}}^{v_A} (M_{J_1}+M_{J_2})
\end{equation}
with
\begin{equation}
  w_{M_{J_1}M_{J_2}}^{v_A} = \sum_{N_AM_{N_A}}
    \int_0^{+\infty} dR_A
    \left( \chi_{M_{J_1}M_{J_2}N_AM_{N_A}}^{v_A}(R_A) \right)^2
\end{equation}
and similarly for B, 3 and 4. In Ref.~\cite{frisch2015} the Er$_2$-Er$_2$ magnetic-dipole interaction energy was calculated using such a two-body expression with experimental values of $d_{v_A}$ and $d_{v_B}$. Indeed evaluating quantitatively each term of Eq.~(\ref{eq:mag-dip}) requires to know in details the nature of the Feshbach states, namely the functions $\chi_{M_{J_1}M_{J_2}N_AM_{N_A}}^{v_A}(R_A)$ and $\chi_{M_{J_3}M_{J_4}N_BM_{N_B}}^{v_B}(R_B)$, which is not possible for the moment in Er$_2$.

As a consequence we consider a simple model, where the two molecules are in the same state $v_A=v_B=v$ imposing a single-channel condition. As each molecule is made of two atoms in the lowest Zeeman sublevel and colliding in $s$ wave, we assume that the resulting molecules are in the $d$-wave bound level, $M_{J_{i,j}}=-J=-6$, $N_A=N_B\equiv N=2$ and $M_{N_A} = M_{N_B}=0$, so that the $R^{-5}$, $R^{-7}$... terms appear in Eq.~(\ref{eq:mag-dip}). The mean interatomic distance inside each molecule is $\langle v_A|R_A|v_A\rangle = \langle v_B|R_B|v_B\rangle = R_0$. We also assume that Eq.~(\ref{eq:mag-dip}) does not couple different molecular states, \textit{i.e.} $v'_A=v_A$ and $v'_B=v_B$, but that it couples different partial waves $L$ and $L'$. The molecules are assumed to collide in the $s$ wave, which is the case in the temperature range of Ref.~\cite{frisch2015}.

Combined with the selection rules (\ref{eq:selec-rule-1})--(\ref{eq:selec-rule-2}) applied for $v'_A=v_A$ and $v'_B=v_B$, the $s$-wave condition imposes that $M'_L=M_L=0$ for all states. Moreover the Clebsch-Gordan coefficients of Eq.~(\ref{eq:mag-dip}) impose $\lambda_A = \lambda_B=0$, 2 and $2N=4$, and so $\lambda = 2$, 4, ..., $4N+2=10$. To evaluate the importance of each term, we compute adiabatic potential-energy curves, obtained after diagonalization of the Hamitonian
\begin{equation}
  \hat{H}_1(R) = \frac{\hbar^2\hat{\vec{L}}{}^2}
                 {2 \mathcal{M}_\mathrm{red} R^2} 
               + \hat{V}_\mathrm{md}(R)
  \label{eq:hamilt-md}
\end{equation}
where $\mathcal{M}_\mathrm{red}$ is a Er$_2$-Er$_2$ reduced mass, and $\hat{\vec{L}}$ the dimensionless angular momentum associated with the collision, and $\hat{V}_\mathrm{md}$ is the magnetic-dipole term given by Eq.~(\ref{eq:mag-dip}), in the basis spanned by $L$ (note that $M_L=0$), for various intermolecular distances $R$.

\begin{figure}
\begin{center}
\includegraphics[width=12cm]{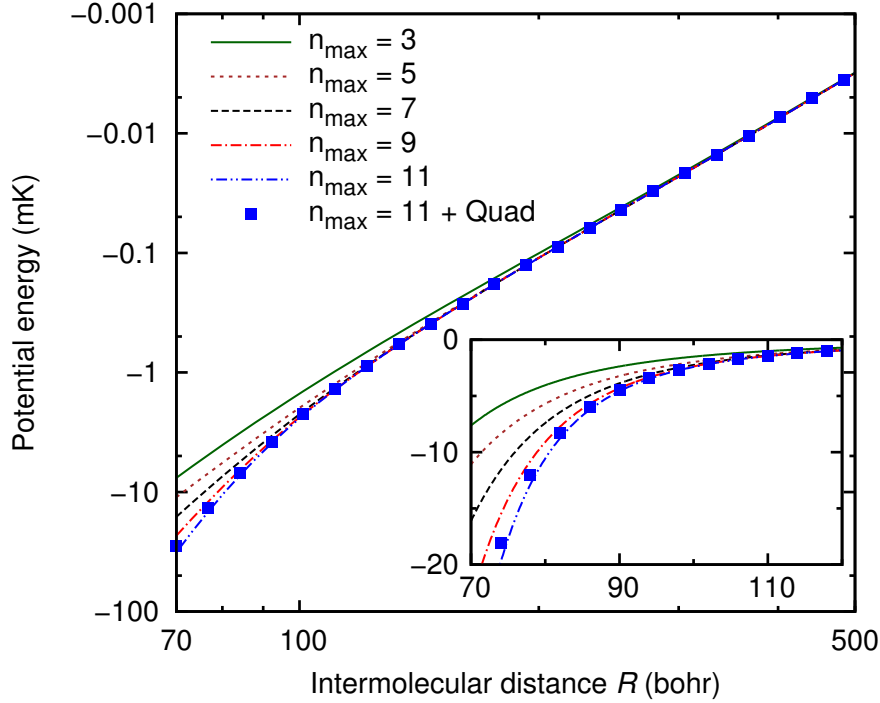}
\caption{\label{fig:PECs-mag-dip} (Color online) Lowest adiabatic potential-energy curves characterizing the magnetic-dipole interaction between two $^{166}$Er$_2$ Feshbach molecules in the same level and colliding in the $s$ wave [see Eqs.~(\ref{eq:mag-dip}) and (\ref{eq:hamilt-md})]. in Eq.~(\ref{eq:mag-dip}) the sum on inverse powers $n=\lambda+1$ of the intermolecular distance $R$ is either stopped at $n_\mathrm{max}=3$ (solid line), $n_\mathrm{max}=5$ (dotted line), $n_\mathrm{max}=7$ (dashed line), $n_\mathrm{max}=9$ (dash-dotted line), or the largest possible value $n_\mathrm{max}=11$ (dash-dot-dotted line). The curves with squares also accounts for the electric-quadrupole interaction (see Eq.~(\ref{eq:hamilt-eq})). The main panel is in log-log scale, and the inset is a linear-scale zoom on the small-distance region. Other parameters are: $R_0=80$ bohr and $L_\mathrm{max}=30$.}
\end{center}
\end{figure}

The results of such calculations are presented on Fig.~\ref{fig:PECs-mag-dip}, when the lowest adiabatic potential-energy curve is plotted in different situations. To highlight the influence of the different $R^{-n}$ terms, we truncate the sum on $\lambda$ in Eq.~(\ref{eq:mag-dip}) up to $\lambda_\mathrm{max}$, which ranges from 2, corresponding to the $R^{-3}$ term, to its largest possible value 10, corresponding to the $R^{-11}$ term. Partial waves from $L=0$ to $L_\mathrm{max}=30$ are included in the basis for a proper convergence. We assume that $\langle R_A^{\lambda_A}\rangle \approx \langle R_A\rangle^{\lambda_A} = R_0^{\lambda_A}$ (and similarly for B), which can be justified by the expected strong localization of the vibrational wave function around the outer classical turning point of the corresponding potential-energy curve. We take $R_0=80$ bohrs as a typical value for $d$-wave resonances observed in \cite{frisch2015}. Strictly speaking our calculation is not valid for $R<R_0$ and so the lower bound of the $x$ axis in Fig.~\ref{fig:PECs-mag-dip} should be $R_0$. But we extend it to a slightly shorter value since $R_0$ is only an estimate.

Figure \ref{fig:PECs-mag-dip} shows that higher-order $R^{-n}$ effective-multipole terms ($n>3$) get more and more important as $R$ decreases. At $R=R_0$ all the terms bring similar contributions to the potential energy. More surprisingly, even at $R=120 \mathrm{\,\, bohrs} = 1.5\times R_0$, the $R^{-3}$ term ($n_\mathrm{max} = 3$) only accounts for two thirds of the total energy. Because the range of energy on Fig.~\ref{fig:PECs-mag-dip}, of a few mK, that is tens of MHz, is larger than the typical energy spacing between neighboring Feshbach levels, couplings with those levels should be taken into account in order to give more accurate predictions.

After the magnetic-dipole interaction, the next term of the multipolar expansion between two erbium atoms is the electric-quadrupole interaction. We have calculated the reduced quadrupole moment of erbium ground level $\langle J=6||\hat{Q}_2||J=6\rangle$ using a Dirac-Hartree-Fock method, and found -1.305 a.u. We have evaluated its impact on the intermolecular interaction by diagonalizing the Hamiltonian
\begin{equation}
  \hat{H}'_1(R) = \frac{\hbar^2\hat{\vec{L}}{}^2}
                  {2 \mathcal{M}_\mathrm{red} R^2} 
                + \hat{V}_\mathrm{md}(R),
                + \hat{V}_\mathrm{eq}(R)
  \label{eq:hamilt-eq}
\end{equation}
where the electric-quadrupole energy $\hat{V}_\mathrm{eq}$ is obtained by setting $\ell_i = \ell_j = 2$, $\ell=4$ and $F_0=1/4\pi\epsilon_0$ in Eq.~(\ref{eq:sum-ij-2}). The latter shows that the electric-quadrupole interaction consists in repulsive contributions scaling from $R^{-5}$ to $R^{-13}$. Figure \ref{fig:PECs-mag-dip} shows that the influence of the quadrupole interaction is visible for $R<90$ bohr, a region where the van der Waals interaction will be actually dominant \cite{kotochigova2011}.

\subsection{Van der Waals interaction \label{sub:vdw}}

The next term of the pairwise atomic multipolar expansion comes from the van der Waals (vdW) interaction, proportional to $R_{ij}^{-6}$.  In Ref.~\cite{lepers2014} we have shown that the Er-Er vdW interaction is mostly isotropic, and characterized by a coefficient $C_{6,000}^\mathrm{Er-Er}=1760$ a.u.. Therefore in this section we calculate the second-order electric-dipole interaction between two weakly-bound diatomic molecules whose atoms interact through an isotropic vdW term. Such a calculation is applicable for Er$_2$-Er$_2$ interactions, but also for molecules made of alkali-metal or alkaline-earth-metal atoms, for which vdW is indeed the strongest interaction.

The vdW energy is a second-order correction due to the electric-dipole interaction, $\ell_i = \ell_j = \ell_{i'} = \ell_{j'}=1$, $\ell=\ell'=2$ and $F_0=1/4\pi\epsilon_0$ in Eq.~(\ref{eq:2ord-oper-3}), while the isotropy results from $k_A = k_B = k=0$. Pointing out that $k=0$ implies $\pi=\kappa$ we obtain for states $v_A=v_B=v$
\begin{eqnarray}
   & & \langle vvL'0|\hat{W}_\mathrm{vdW}|vvL0\rangle
  \nonumber \\
   & = & -\frac{C_{6,000}^\mathrm{Er-Er}}
               {6(4\pi\varepsilon_0)^2R^6}
    \sum_{\lambda,\lambda'=2}^{+\infty}
    \sum_{\lambda_A,\lambda_B=0,2,...}^{\lambda}
    \sum_{\lambda'_A,\lambda'_B=0,2,...}^{\lambda'}
    \delta_{\lambda_A+\lambda_B+2,\lambda}
    \delta_{\lambda'_A+\lambda'_B+2,\lambda'}
    \frac{\langle R_A^{\lambda_A+\lambda'_A}\rangle
          \langle R_B^{\lambda_B+\lambda'_B}\rangle}
         {(2R)^{\lambda_A+\lambda_B+\lambda'_A+\lambda'_B}}
  \nonumber \\
   & \times & \sqrt{\frac{(2\lambda+1)!(2\lambda'+1)!}
                         {(2\lambda_A)!(2\lambda_B)!
                          (2\lambda'_A)!(2\lambda'_B)!}}
    \sqrt{(2\lambda-3)(2\lambda'-3)(2\kappa_A+1)(2\kappa_B+1)}
  \nonumber \\
   & \times & \sum_{\kappa_A,\kappa_B,\kappa}
    \ninej{\lambda_A}{\lambda_B}{\lambda_A+\lambda_B}
          {\lambda'_A}{\lambda'_B}{\lambda'_A+\lambda'_B}
          {\kappa_A}{\kappa_B}{\kappa}
    \sixj{\lambda_A+\lambda_B}{\lambda'_A+\lambda'_B}{\kappa}
         {\lambda'}{\lambda}{2}
  \nonumber \\
   & \times & \mathcal{C}_{\lambda_A0\lambda'_A0}^{\kappa_A0}
    \mathcal{C}_{\lambda_B0\lambda'_B0}^{\kappa_B0}
    \mathcal{C}_{\lambda0\lambda'0}^{\kappa0}
    \mathcal{C}_{\kappa_A0\kappa_B0}^{\kappa0}
    \sqrt{\frac{2L'+1}{2L+1}} \left( 
    \mathcal{C}_{L'0\kappa0}^{L0} \mathcal{C}_{N0\kappa_A0}^{N0}
    \mathcal{C}_{N0\kappa_B0}^{N0}  \right)^2
  \label{eq:vdw}
\end{eqnarray}
where we used
\begin{equation}
  C_{6,000}^\mathrm{Er-Er} \equiv C_{6,000}^{i,j}
    = \frac{1}{2} \sum'_{i'',j''}
    \frac{\mathcal{Q}_{(\ell_i=1,\ell_i=1)00}''
          \mathcal{Q}_{(\ell_j=1,\ell_j=1)00}''}
         {\Delta_{i''}+\Delta_{j''}}
\end{equation}
and
\begin{equation}
  \ninej{a}{b}{c}{d}{e}{f}{g}{h}{0} =
    \frac{(-1)^{b+c+d+g}\delta_{cf}\delta_{gh}}
         {\sqrt{(2c+1)(2g+1)}} \sixj{a}{b}{c}{e}{d}{g} ,
\end{equation}
with $\{:::\}$ a Wigner 6-j symbol, and where $\langle R_{A,B}^{\lambda_{A,B}+\lambda'_{A,B}}\rangle$ is the interatomic distance at the corresponding power in molecules A and B, averaged over the vibrational wave function of state $|v\rangle$.
Equation (\ref{eq:vdw}) is a series of terms proportional to $R^{-n}$. The leading term, which scales as $R^{-6}$, comes out when $\lambda_A = \lambda_B = \lambda'_A = \lambda'_B = 0$, and so $\kappa_A = \kappa_B = \kappa = 0$. It is thus a fully isotropic term ($L=L'$) equal to $-4C_{6,000}^\mathrm{Er-Er} / R^6$. The next terms scale as $R^{-8}$, $R^{-10}$, .... Unlike the first-order expression (\ref{eq:mag-dip}), $n$ goes \textit{a priori} to infinity, since it is not limited by the angular selection rules. The bounds in the sums over $\kappa_A$, $\kappa_B$ and $\kappa$ are not explicitly specified, as they come from several conditions. The 9-j symbol of Eq.~(\ref{eq:vdw}) imposes $|\lambda_A-\lambda'_A|\le\kappa_A\le \lambda_A+\lambda'_A$, whereas the Clebsch-Gordan coefficient of the last line imposes $0\le \kappa_A \le 2N$ and $\kappa_A$ even. The most restrictive conditions will indeed apply, namely $\max( |\lambda_A-\lambda'_A|,0) \le  \kappa_A \le \min( \lambda_A+\lambda'_A,2N)$, $\kappa_A$ even. The conditions are similar for $\kappa_B$, while for $\kappa$ we have: $|\lambda_A+\lambda_B-\lambda'_A-\lambda'_B| \le\kappa \le (\lambda_A+\lambda_B+\lambda'_A+\lambda'_B)$, $|\kappa_A-\kappa_B| \le\kappa \le (\kappa_A+\kappa_B)$, $|L-L'|\le \kappa \le (L+L')$, and $\kappa$ even.

\begin{figure}
\begin{center}
\includegraphics[width=12cm]{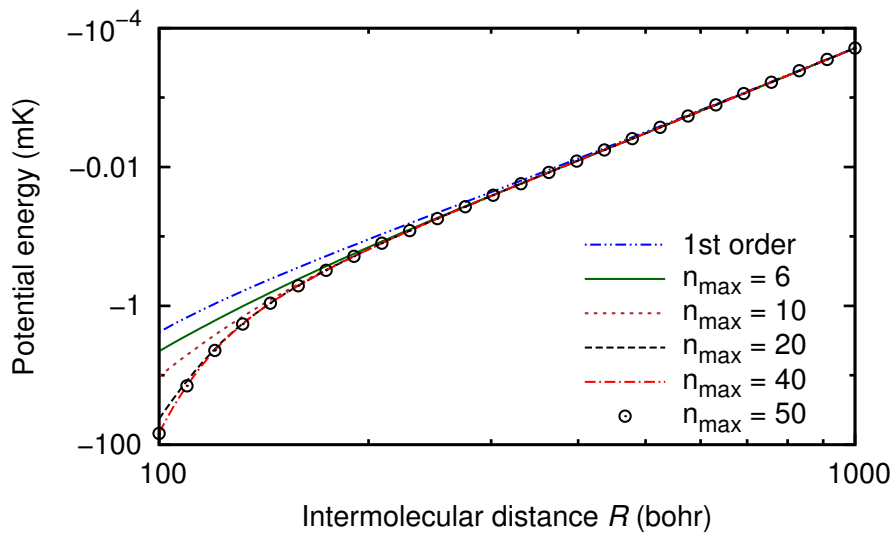}
\caption{\label{fig:PECs-vdw} (Color online) Lowest adiabatic potential-energy curves characterizing the magnetic-dipole, electric-quadrupole and van der Waals interactions between two $^{166}$Er$_2$ Feshbach molecules in the same level and colliding in $s$ wave.
In Eq.~(\ref{eq:vdw}) the sum on $\lambda_A$, $\lambda_B$, $\lambda'_A$ and $\lambda'_B$ is stopped at different values of $n_\mathrm{max}$, where $n=6 + \lambda_A + \lambda_B + \lambda'_A + \lambda'_B$, namely $n_\mathrm{max}=6$ (solid line), $n_\mathrm{max}=10$ (dotted line), $n_\mathrm{max}=20$ (dashed line), $\lambda_\mathrm{max}=40$ (dash-dotted line) and $n_\mathrm{max}=50$ (open circles). For comparison the lowest adiabatic PEC only accounting for the first-order magnetic-dipole and electric-quadrupole interactions is shown again (dash-dot-dotted line). Other parameters: $R_0=80$ bohr and $L_\mathrm{max}=30$.
}
\end{center}
\end{figure}

The convergence of the $R^{-n}$ series in Eq.~(\ref{eq:vdw}) is addressed on figure \ref{fig:PECs-vdw}, where we plot the lowest adiabatic PEC obtained after diagonalization of the hamiltonian $\hat{H}_2(R) = \hat{H}'_1(R) + \hat{W}_\mathrm{vdW}(R)$ [see Eqs.~(\ref{eq:hamilt-eq}) and (\ref{eq:vdw})] including the magnetic-dipole, electric-quadrupole and van der Waals interactions, and where again we assume $\langle R_A^{\lambda_A+\lambda'_A}\rangle \langle R_B^{\lambda_B+\lambda'_B}\rangle \approx R_0^{\lambda_A+\lambda'_A+\lambda_B+\lambda'_B}$. Contrary to the first-order correction, convergence will require higher terms as the ratio $R_0/R$ decrease. At $R=100$ bohr, that is to say $R_0/R=4/5$, convergence is reached for $n_\mathrm{max}=40$, whereas for $R<R_0$, the condition $r''<r'$ to apply Eq.~(\ref{eq:Ckq-1}) indicates that convergence cannot be achieved.
In addition, we see that the van der Waals energy is significantly larger than the first-order energies on the left part of Fig.~\ref{fig:PECs-vdw}; at $R=120$ bohr it represents 73 \% of the total potential energy.

\subsection{Ultracold collisions and characteristic lengths}

In order to estimate the role played in collisions at ultralow energies by the additional terms of the multipolar expansion, it is instructive to calculate the so-called characteristic length associated with each of those terms \cite{julienne1989, williams1999}. In our case, a given term will cause significant reflection of the incoming scattering wave function, if its characteristic length is larger than the typical size $R_0$ of each molecule.

We start with discussing the influence of the magnetic-dipole interaction. We assume that  each $R^{-n}$ term of Eq.~(\ref{eq:mag-dip}) is described by a $C_n$ coefficient
\begin{equation}
  C_n \approx \frac{\mu_0}{4\pi}
              \left(2\mu_B g_J J\right)^2 R_0^{n-3}
      = C_3 R_0^{n-3} \,,
  \label{eq:Cn}
\end{equation}
where we take the largest possible magnetic moments $d_{v_A} = d_{v_B} = -2\mu_B g_J J$ for molecules A and B. The characteristic length associated with the term $C_n/R^n$ is \cite{julienne1989, williams1999}
\begin{equation}
  a_n = \frac{1}{2} \left(
           \frac{2\mathcal{M}_\mathrm{red}C_n}
                {\hbar^2}\right)^{\frac{1}{n-2}}.
  \label{eq:an}
\end{equation}
By inserting (\ref{eq:Cn}) into (\ref{eq:an}), we can express all the characteristic lengths as functions of $a_3=\mathcal{M}_\mathrm{red}C_3 / \hbar^2$
\begin{equation}
  \left( \frac{a_n}{R_0} \right) =
    \left( \frac{a_3}{R_0} \right) ^{\frac{1}{n-2}},
  \label{eq:an-a3}
\end{equation}
which gives $(a_5/R_0)= (a_3/R_0)^{1/3}$, $(a_7/R_0)= (a_3/R_0)^{1/5}$, and so on.

Equation (\ref{eq:an-a3}) shows that if $a_3> R_0$, then $a_3> a_5> a_7> ...> R_0$, for all $n$, and \textit{vice versa}. So if the $R^{-3}$ term significantly influences the ultracold dynamics, the $R^{-5}$, $R^{-7}$, ... terms associated with effective molecular multipole moments of higher ranks are also likely to do so. This is indeed the case in our present Er$_2$-Er$_2$ study where $R_0=80$~bohr, $a_3=790$~bohr, $a_5=172$~bohr, $a_7=126$~bohr, etc. This reasoning can be generalized to all pairwise atomic interactions. For a {}``parent" term scaling as $C_p/R^p$ and associated with the characteristic length $a_p$ (see Eq.~(\ref{eq:an})), the related term $C_n/R^n = R_0^{n-p}C_p/R^n$ is associated with the characteristic length 
\begin{equation}
  \left( \frac{a_n}{R_0} \right) =
    \left( \frac{a_p}{R_0} \right) ^{\frac{p-2}{n-2}},
  \label{eq:an-ap}
\end{equation}
for $n\ge p$.
Considering the van der Waals interaction $p=6$ with $C_6 = 4\times C_{6,000}^\mathrm{Er-Er} = 7040$~a.u., we obtain $a_6 = 128$~a.u., $a_8 = 109$~a.u., $a_{10} = 101$~a.u., etc. The terms coming from the van der Waals interactions are also likely to play a crucial role.

\begin{figure}
\begin{center}
\includegraphics[width=12cm]{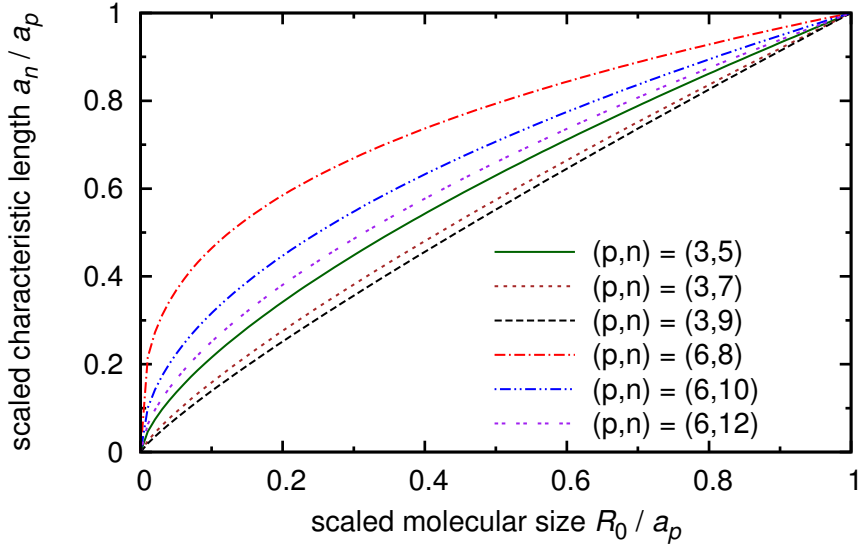}
\caption{\label{fig:charac-len} (Color online) Characteristic length $a_n$ as function of the size $R_0$ of the molecules. Both quantities are rescaled with respect to their {}``parent" characteristic length, namely $a_{p=3}$ for magnetic-dipole and $a_{p=6}$ for van der Waals interaction (see Eq.~(\ref{eq:an-ap})).
}
\end{center}
\end{figure}

Finally we investigate the influence of the size of individual molecules, which, for halo molecular states close to a Feshbach resonance, is equivalent to the atom-atom scattering length \cite{chin2010}. To that end we put the $R_0$ term in the right-hand side of Eq.~(\ref{eq:an-ap}) and rescale both sides with respect to $a_p$
\begin{equation}
  \left( \frac{a_n}{a_p} \right) =
    \left( \frac{R_0}{a_p} \right) ^{\frac{n-p}{n-2}}.
  \label{eq:an-R0}
\end{equation}
Figure \ref{fig:charac-len} shows this quantity as function of the scaled molecular size, for the three terms $n=p+2$, $p+4$ and $p+6$ associated with the magnetic-dipole $p=3$ and van der Waals  interactions $p=6$. When the size of the molecule is much smaller than the parent characteristic length ($R_0\ll a_p$), the scaled characteristic lengths $a_n/a_p$ vanish and the important interactions are the dipole-dipole interaction $p=3$ and the van der Waals interaction $p=6$. But when the molecular size increases up to the parent characteristic length ($R_0\approx a_p$), $a_n\to a_p$ and all the higher molecular effective multipole moments become significant in the overall interaction and hence in the ultracold dynamics.

\section{Concluding remarks \label{sec:conclu}}

In this article, we present the general formalism to characterize the long-range interactions between two arbitrary charge distributions, each composed of almost free subsystems, which we apply to the case of weakly-bound diatomic molecules. To that end, considering that the atoms composing each molecule conserve their identity, we expand the intermolecular potential energy as the sum of pairwise atomic energies. By expressing the intermolecular potential energy as a function of the intermolecular distance, we obtain a generalization of the usual multipolar expansion, containing additional terms scaling as inverse powers of the intermolecular distance. Those additional terms, which involve effective molecular multipole moments, are strongly anisotropic with respect to the molecular orientations.

In the case of two vibrationally highly-excited Er$_2$ molecules, many additional terms bring a substantial contribution to the intermolecular potential energy. By estimating their characteristic lengths, we predict that those additional terms are also likely to influence the Er$_2$-Er$_2$ collisions at ultralow energies. To confirm that prediction, we can perform quantum-scattering calculations using the intermolecular potential-energy curves presented in this article. This would require however to know precisely the multi-channel wave function of the Feshbach states. Besides, since the intermolecular potential energy is a few mK (see Fig.\ref{fig:PECs-vdw}), which is the typical spacing between neighboring Feshbach levels, we can expect the intermolecular interaction to couple different Feshbach levels, and so to induce inelastic collisions.

\section*{Acknowledgements}

The authors acknowledge support from {}``Agence Nationale de la Recherche'' (ANR), under the project COPOMOL (contract ANR-13-IS04-0004-01), and from Universit\'e Paris-Sud under the project {}``Attractivit\'e 2014".

\appendix

\section{Second-order correction and irreducible tensors \label{sec:app}}

In order to write Eq.~(\ref{eq:2ord-oper-2}) as a sum of irreducible tensors, firstly we expand the product of atomic multipole moments as
\begin{equation}
  Q_{\ell_{i}m_{i}} |1''2''\rangle
  \langle 1''2''| Q_{\ell_{i'}m_{i'}} = 
    \sum_{k_A=|\ell_i-\ell_{i'}|}
        ^{\ell_i+\ell_{i'}}
    \sum_{q_A=-k_A}^{+k_A}
      \mathcal{C}_{\ell_{i}m_{i}\ell_{i'}m_{i'}}^{k_A q_A}
      \mathcal{Q}_{(\ell_i,\ell_{i'})k_A q_A}''
  \label{eq:cpl-mult-inv}
\end{equation}
where $\mathcal{Q}_{(\ell_i,\ell_{i'})k_A q_A}''$ are the coupled atomic multipole moments (see Eq.~(\ref{eq:cpl-mult})), and similarly for atoms $j$ and $j'$ of molecule B. Then we apply the transformation (see Ref.~\cite{varshalovich1988}, Ch.~8, Eq.~(20), p.~260)
\begin{equation}
  \sum_{\beta\gamma\varepsilon\varphi}
    \mathcal{C}_{b\beta c\gamma}^{a\alpha}
    \mathcal{C}_{e\varepsilon f\varphi}^{d\delta}
    \mathcal{C}_{e\varepsilon b\beta}^{g\eta}
    \mathcal{C}_{f\varphi c\gamma}^{j\mu}=[adgj]^{1/2}
    \sum_{t\tau} \mathcal{C}_{g\eta j\mu}^{t\tau}
    \mathcal{C}_{d\delta a\alpha}^{t\tau}
    \ninej{c}{b}{a}{f}{e}{d}{j}{g}{t} ,
  \label{eq:4cg}
\end{equation}
where $[x_1x_2...x_n]=(2x_1+1)(2x_2+1)\times ...\times(2x_n+1)$ and the number between curly brackets is a Wigner 9-j symbol, to $a=\ell'$, $b=\ell_{i'}$, $c=\ell_{j'}$, $d=\ell$, $e=\ell_i$, $f=\ell_j$, $g=k_A$, $j=k_B$ and $t=k$ and to the corresponding components. We used that a 9-j symbol is unchanged after a permutation of two rows followed by a permutation of two columns.

Secondly we work out the effective molecular multipole moments. We expand the products of Racah spherical harmonics in Eq.~(\ref{eq:g-ij-3}) as (see Ref.~\cite{varshalovich1988}, Ch.~5, Eq.~(9), p.~144)
\begin{eqnarray}
  C_{\lambda_{i }\mu_{i }}(\Theta_{i },\Phi_{i })
  C_{\lambda_{i'}\mu_{i'}}(\Theta_{i'},\Phi_{i'})
   & = & (-1)^{\delta_{i2}\lambda_i
              +\delta_{i'2}\lambda_{i'}} \,
  C_{\lambda_{i }\mu_{i }}(\Theta_A,\Phi_A)
  C_{\lambda_{i'}\mu_{i'}}(\Theta_A,\Phi_A)
  \nonumber \\
   & = & (-1)^{\delta_{i2}\lambda_i
              +\delta_{i'2}\lambda_{i'}}
  \sum_{\kappa_A=|\lambda_i-\lambda_{i'}|}
      ^{\lambda_i+\lambda_{i'}}
  \sum_{\varrho_A=-\kappa_A}^{+\kappa_A}
  \mathcal{C}_{\lambda_i\mu_i\lambda_{i'}\mu_{i'}}
             ^{\kappa_A\varrho_A}
  \nonumber \\
   & & \times C_{\kappa_A\varrho_A}
               (\Theta_A,\Phi_A)
  \mathcal{C}_{\lambda_i0\lambda_{i'}0}^{\kappa_A0}
  \label{eq:Ccpl}
\end{eqnarray}
where, recalling that $\Theta_A = \Theta_1 = \pi-\Theta_2$ and $\Phi_A = \Phi_1 = \pi+\Phi_2$, we used the property $C_{\lambda_2\mu_2}(\Theta_2,\Phi_2) = (-1)^{\lambda_2}\,C_{\lambda_2\mu_2}(\Theta_A,\Phi_A)$. After writing a similar equation for molecule B, we apply again Eq.~(\ref{eq:4cg}) to $a=\lambda'-\ell_{i'j'}=\lambda_{i'}+\lambda_{j'}$, $b=\lambda_{i'}$, $c=\lambda_{j'}$, $d=\lambda-\ell_{ij}=\lambda_i+\lambda_j$,  $e=\lambda_i$, $f=\lambda_j$, $g=\kappa_A$, $j=\kappa_B$, $t=\pi$, and the corresponding components including $\tau=\sigma$. Then we apply the formula (see Ref.~\cite{varshalovich1988}, Ch.~8, Eq.~(26), p.~261)
\begin{equation}
  \sum_{\beta\gamma\varepsilon\varphi}
    \mathcal{C}_{b\beta c\gamma}^{a\alpha}
    \mathcal{C}_{e\varepsilon f\varphi}^{d\delta}
    \mathcal{C}_{e\varepsilon g\eta}^{b\beta}
    \mathcal{C}_{f\varphi j\mu}^{c\gamma}=[bcdt]^{1/2}
    \sum_{t\tau} \mathcal{C}_{g\eta j\mu}^{t\tau}
    \mathcal{C}_{d\delta t\tau}^{a\alpha}
    \ninej{a}{b}{c}{d}{e}{f}{t}{g}{j} ,
  \label{eq:4cg-2}
\end{equation}
to $a=k$, $b=\ell_{ij}$, $c=\ell_{i'j'}$, $d=\pi$, $e=\lambda_i+\lambda_j$, $f=\lambda_{i'}+\lambda_{j'}$, $g=\lambda$, $j=\lambda'$, $t=\kappa$, and the corresponding components including $\tau=\varrho$. Then the invariance of the 9-j symbol: (i) by reflection about the anti-diagonal; (ii) by permutation of the resulting first two lines and the first two columns. Afterwards we deal with the Racah spherical harmonics for the intermolecular axis. Applying (see Ref.~\cite{varshalovich1988}, Ch.~5, Eq.~(10), p.~144)
\begin{equation}
  \sum_{\alpha\beta} \mathcal{C}_{a\alpha b\beta}^{c\gamma} C_{a\alpha}(\theta,\phi) C_{b\beta}(\theta,\phi)
   = \mathcal{C}_{a0b0}^{c0} C_{c\gamma}(\theta,\phi)
\end{equation}  
to $a=\kappa_A$, $b=\kappa_B$, $c=\kappa$, $\alpha=\varrho_A$, $\beta=\varrho_B$ and $\gamma=\varrho$, we get to Eq.~(\ref{eq:2ord-oper-3}).

\section*{Bibliography}


\end{document}